\documentclass[letterpaper, 10 pt, conference]{ieeeconf}  % Comment this line out if you need a4paper

\IEEEoverridecommandlockouts                              % This command is only needed if 
\usepackage[noadjust]{cite}
\title{\LARGE \bf
Providing a Philosophical Critique and Guidance of Fairness Metrics
}

\author{Henry Cerbone$^{1}$% <-this % stops a space
\thanks{$^{1}$ Harvard College, hcerbone@college.harvard.edu}}

\begin{document}

\maketitle
\thispagestyle{empty}
\pagestyle{empty}

%%%%%%%%%%%%%%%%%%%%%%%%%%%%%%%%%%%%%%%%%%%%%%%%%%%%%%%%%%%%%%%%%%%%%%%%%%%%%%%%
\begin{abstract}

 In this project, I seek to present a summarization and unpacking of themes of fairness both in the field of computer science and philosophy.  This is motivated by an increased dependence on notions of fairness in computer science and the millennia of thought on the subject in the field of philosophy.  It is my hope that this acts as a crash course in \emph{fairness philosophy} for the everyday computer scientist and specifically roboticist. This paper will consider current state-of-the-art ideas in computer science, specifically algorithmic fairness, as well as attempt to lay out a  rough set of guidelines for metric fairness .  Throughout the discussion of philosophy, we will return to a thought experiment posed by Cynthia Dwork on the question of randomness.

\end{abstract}

%%%%%%%%%%%%%%%%%%%%%%%%%%%%%%%%%%%%%%%%%%%%%%%%%%%%%%%%%%%%%%%%%%%%%%%%%%%%%%%%
\section{INTRODUCTION}
As roboticists, we are becoming increasingly concerned with whether the systems we create are fair. Corresponding to this concern, there is a pressure to figure out just what it means to be fair. Recent work such as \cite{fairrobotics} has tended to reference prevailing work by Dwork et. al. \cite{fta} in computer science on algorithmic fairness. In doing so, researchers inherit the implications for pure metric fairness regardless of its later application in the particular robotics context. In this paper, I set out what exactly that baggage looks like. To do this, work in philosophy is drawn out to lay out not only what definition of fairness the prevailing technical definition relies on but also how it \emph{should} look. Unlike work such as \cite{fairnessdefinitions}, this is not a survey or ordering of definitions of fairness. I open with the motivation for pursuing a philosophical critique and guidance to algorithmic fairness before outlining what the views of fairness are from each field. I close with how to marry these two views in particular what can be learned from philosophy. This work serves not as a solution from philosophy for fairness in computer science, but instead, as a rough road map for researchers hoping to better understand the work surrounding what it means to be fair. Additionally, it is useful to inherit the definition of robot from Coglianese et. al. \cite{CoglianeseCary2017RbrA} which defines an algorithm as a ``digital robot". Under this definition, we are able to both directly apply the discussion of fairness from computer science to robotics as well as better separate the component parts of a system and reason about them philosophically. 

\section{MOTIVATION}

We find ourselves setting out in our design and creation of algorithms with a sense of duty.  This is sourced from the increasing reality that our algorithms' decisions affect real peoples' lives in real ways. One result of this duty is that we should take into account how these algorithms make these decisions.  In modern society, we are increasingly caught up in questions of what is fair or what is just.  This is largely a result of a basic human drive for fairness on a personal level and a reaction to centuries of oppression of people based on race, gender, and social class.  As computer scientists, we have also begun an investigation into what is fair and what is not.  This is a result of the growing field of algorithmic fairness and people designing algorithms of the aforementioned sort.  In our investigation of fairness, it is important to remember the words of Ludwig Wittgenstein in his \textit{Philosophical Investigations} \cite{WittgensteinLudwig2001Pi:t}. 
\begin{quote}
 We have got on to slippery ice where there is no friction, and so, in a certain sense, the conditions are ideal; but also, just because of that, we are unable to walk.  We want to walk: so we need friction.  Back to the rough ground!
\end{quote}
In other words, we are asking what is fair for real people living real lives.  A definition or notion of fairness that is not contextualized or is overly abstracted will fail to capture its sole benefactor: people.
We are motivated in our investigation of what is fair to turn to a field that has produced the greatest volume of material on the topic --- philosophy. It is with the hope of applying some of the philosophical thought of the last century to new areas of computer science that we arrive at something of use to researchers working in algorithmic fairness and computer science/robotics more broadly.

\section{Perspective: Algorithmic Fairness}

The primary notion of fairness in computer science is from the definition given by Dwork et. al. in \cite{fta} as given below
\begin{quote}
 We capture fairness by the principle that any two individuals who are similar with respect to a particular task should be classified similarly. In order to accomplish this individual-based fairness, we assume a distance metric that defines the similarity between the individuals. This is the source of “awareness” in the title of this paper. We formalize this guiding principle as a Lipschitz condition on the classifier. In our approach a classifier is a randomized mapping from individuals to outcomes, or equivalently, a mapping from individuals to distributions over outcomes. The Lipschitz condition requires that any two individuals $x, y$ that are at distance $d(x, y) \in [0, 1]$ map to distributions $M(x)$ and $M(y)$, respectively, such that the statistical distance between $M(x)$ and $M(y)$ is at most $d(x, y)$. In other words, the distributions over outcomes observed by x and y are indistinguishable up to their distance $d(x, y)$.
\end{quote}
This definition is certainly appealing.  It is intuitive that in a discussion of fairness, we want people that are similar to be treated in a similar fashion.  The obvious question that arises from this definition, however, is --- What does it mean to be similar?  In the context of the paper, how do we come up with a metric?  As mentioned in the paper, it seems that the goal of the metric definition is to have a Rawlsian style mechanism of social revision.  This is important as otherwise we seem to have a circular notion with a distinction of good similarities and bad similarities, important ones and non-important ones.  The goal of imposing the metric externally is certainly sound.  An alternative is given by Ilvento in \cite{ilvento} through the inclusion of a human arbiter.  The potential problems with these approaches arise from a closer examination of the philosophy on which they are based.

\section{Perspective: Philosophy}
\subsection{\indent A Question of Randomness}
Throughout our discussion of the philosophical perspective, we will find it useful to return to a concrete example, to stay on the rough ground so to speak.  We will rely on an example from Professor Cynthia Dwork which was brought out in conversation with Deborah Hellman.
\begin{quote}
 Suppose at interview time the interviewer were to flip a coin to choose a letter of the alphabet and to summarily reject all applications with names beginning with the chosen letter.  This seems “fair” but arbitrary, regardless of how it feels to Allen and Anna.  If instead the interviewer has a fixed antipathy towards “A”, can we view this as having the randomness occurring 20 years earlier, at the time the baby was named?  Does the timing of the randomness matter?
\end{quote}
and
\begin{quote}
 Suppose we have 10 sentencing cases and 5 prison beds.  Apparently, it is unacceptable to randomly choose a subset of the 5 of the felons for incarceration.  But it is viewed as OK to randomly order the cases, and for the judge to send the first 5 to prison and release the next 5 due to overcrowding.
\end{quote}
I believe that these examples allow us to think through the various advantages and disadvantages brought to the table by various philosophical views.  They also contain simplified pieces of key aspects of algorithmic fairness.  Namely, the first example deals with randomness which we often think of as a cure-all for fair distribution.  The second example focuses on the problem of contextualization and implicit effects.

\subsection{\indent Rawls}
Although the discussion of “What is it to be fair?” is a discussion that has persisted for millennia in philosophy, we will take the work of John Rawls as a starting point for the discussion of fairness in philosophy.  Rawls’ book, \emph{Theory of Justice} \cite{rawls}, was a groundbreaking work in political philosophy.  We will borrow several small parts from this work in our discussion but will not discuss it in detail.  Also, of particular interest to us is a paper published as a precursor to \emph{Theory of Justice} entitled \emph{Justice as Fairness} \cite{rawls2}.
\subsubsection{\indent\indent Justice as Fairness}
Rawls gives an account of justice that takes itself as relying on fairness. Much like notions in computer science, one worries that this is shifting the question. However, in \cite{rawls2}, Rawls gives us a thought experiment in which he gives us a picture of what fairness might be.  If we imagine members of a society coming together to select the rules on which they should be judged, knowing that these rules once agreed upon will be used to judge all other members, this must be regarded as fair.  This is an interesting and attractive notion of what is fair.  It seems to be simple and disposes with what is known as the veil of ignorance from Rawls’ later work.  We can also see how this has direct consequences for metric fairness.  If people who are being assigned positions by an algorithm all mutually agree upon the metric with which they will be judged, it seems to follow that they have a prima facia right to follow the rulings.
\subsubsection{\indent\indent The Veil of Ignorance and The Difference Principle}
The primary takeaways from \cite{rawls} for this context are the veil of ignorance and the difference principle.  The veil of ignorance is Rawls’ requirement that in order to be fair/just, we must imagine ourselves behind a veil, ignorant of our own standing and selves.  Behind this veil, we are all rational, moral beings with equal standing.  This differs from the aforementioned system in which people make rules with knowledge of how they themselves will be affected.  It also has interesting implications for computer science. Namely, we can imagine an algorithm as learning metric fairness such as the methods found by Gillen et. al in \cite{gillen}. This is primarily due to algorithms operating behind a veil of ignorance as they obviously have no self to be concerned about. Although this claim cannot be taken to be self-evident, various scenarios of machine learning based solutions built upon many scenarios of fairness across all levels of society would then be operating behind a veil of ignorance (or at least more ignorance than Rawls' law makers). Behind this veil of ignorance, Rawls’ believes we must maintain two principles, the latter of which is the difference principle.  The difference principle states that social and economic inequalities must satisfy that they are the greatest benefit to the least advantaged \cite{rawls}. We see here a departure from both ``similar people are treated similarly” and from a random approach as previously presented.  What ensues can be seen as a balancing act between differentials and allowing disadvantaged equal access to social and political offices.
\smallbreak
The importance of the Rawlsian perspective of fairness to this paper is most obviously in its citation in algorithmic fairness literature. Computer scientists read Rawls as supporting the notion of metric fairness presented earlier. Whether or not this is an accurate reading is out of the scope of this work, however, egalitarianism as a whole is both influenced and in some ways responding to Rawls. Therefore, presenting Rawls' views are essential to any discussion of fairness. The trend toward a more complete Rawlsian approach to fairness as discussed is starting to be seen such as in work by Hurtado et. al. \cite{Hurtado}. This work appropriately leveraged a relearning step to fix initial preconceptions of what was fair. This is similar to Rawls' principle of reflective equilibrium which serves to correct for decisions initially made behind the veil of ignorance.
\subsection{\indent Egalitarianism}
It is natural to turn to egalitarianism as the next step in our discussion of fairness in the field of philosophy.  This is a natural point of interest as the core premise of egalitarianism is that people are equal and deserve rights and opportunities.  Of particular interest is the debate between luck and relational egalitarians.  This debate begin with the 1999 paper by Elizabeth Anderson entitled \emph{What is the point of equality?}. We will examine both sides both by presenting them and using their methods to work through our randomness thought experiments.
\subsubsection{\indent\indent Luck Egalitarianism}
The basic premise of luck egalitarianism is that any inequality due to morally arbitrary circumstances is unjust.  The goal then is at the end of the day to balance out these lucky advantages to reach equality.  A primary figure in luck egalitarianism, John Roemer, gives us the following review of how one might equalize these inequalities/what should be viewed as such in \cite{roemer}
\begin{quote}
 John Rawls proposes ``primary goods"; Ronald Dworkin proposes ``resources," a re-source being defined in a comprehensive way to include various talents and handicaps; Amartya Sen proposes ``capabilities to function"; in two recent articles, Richard Arneson has proposed ``opportunity for welfare" as the appropriate equalisandum; G. A. Cohen has recently proposed ``access to advantage." All of these proposals attempt to equalize opportunities, rather than outcomes: for Rawls and Dworkin, primary goods and resources, respectively, are the wherewithal with which people carry out projects that lead to outcomes that have value to them; for Sen, the capabilities to function in various ways are the prerequisites for what individuals make of themselves; and Cohen's ``access" is similar to Arne- son's straightforward ``opportunity."
\end{quote}
Returning to our examples, we see that per a luck egalitarian our first example is unjust.  We are saying that people with a first name beginning with \emph{A}, a trait that is morally arbitrary, are to be treated differently.  Our second example gives an interesting distinction to luck egalitarianism.  We see why the two posed scenarios are different.  In the first, everyone is treated justly as we have a random subset.  In the second, we have some people being treated unjustly as a morally arbitrary circumstance has sentenced them a fate conditioned by overcrowding.
If the luck egalitarian breakdown of the second example leaves a bad taste in our mouth, this is a natural reaction.  Luck egalitarianism suffers from a seemingly absurd distinction. The clearest example of this is pointed out by Elizabeth Anderson in \cite{ea},  with the following example from G. Cohen
\begin{quote}
 A talented person (call her Sue) who prefers gardening at the average wage to being a doctor at the same wage should be motivated by the egalitarian ethos to be a doctor because this benefits the least advantaged more than being a gardener (Cohen 2008, 184-85).
 In fact, the difference principle does not give anyone the aim of maximizing benefit to the least advantaged. It only constrains inequality. If Sue chose gardening, she would not make more than others, so she would create no inequality that would have to be justified as maximizing the good of the least well-off. Cohen suggests that Sue, in
 choosing to garden, has created an inequality, since, as one who loves to garden, she enjoys higher work satisfaction than the least advantaged.
\end{quote}
We see here how many views of luck egalitarians seem to offer absurd, inhuman results when applied to seemingly basic scenarios.  It is this lack of intuitiveness that gives rise to the subfield termed relational egalitarianism.
\subsubsection{\indent\indent Relational Egalitarianism}
As previously mentioned, the obvious problems of luck egalitarianism give rise to relational egalitarianism.  Participants in this field take themselves as relying on some deeper moral notion of fairness to judge equality and fairness. This is best illustrated in the coiner of the term \emph{relational egalitarian}, Elizabeth Anderson's closing lines of her paper \emph{The Fundamental Disagreement between Luck and Relational Egalitarians}
\begin{quote}
 Once everyone has done everything justice requires of them, the world is just, whatever other negative evaluations one might make of it. For justice is fundamentally a virtue of agents, not a distributive pattern.
\end{quote}
There is not as rigorous or fleshed out a theory amongst relational egalitarians as to what fairness is. Looking at Anderson's example from above, we see that what is required technically is something more than an ML classification. It is not sufficient to group a number of cases together, label them as fair, and proceed. In a similar vein, it is also not sufficient per relational egalitarianism to say that ``similar individuals are treated similarly". This is because we are creating a distributive pattern for fairness which empties out the meaning of fairness. What Anderson encourages us to do is to allow what is fair (justice in this case) to be the result of the virtue of agents, a result of a moral standing not a statistical one.   
\smallbreak
A modern relational egalitarian (or something akin to such) can be found in \cite{hellman} by Deborah Hellman. Although a full discussion of her book is out of the scope of this work, her notion of discrimination as compounding previous wrongs is a good example of a relational egalitarian standpoint. In this view, we can return to our examples and see that in both scenarios, we cannot say that we have witnessed a case of injustice or discrimination.

\section{Directions for Robotics}
Now that we have established various perspectives of what it means to be fair, we are at a crossroads as to what to do next. The field of fairness in robotics is still in its infancy and therefore, a review of technical results here would be short lived. It is worthwhile to note that work by both Brandao et. al. \cite{BRANDAO2020103259} and Claure et. al. \cite{claure} provide initial technical applications of various notions of fairness. What this work provides additionally is a comprehensive explication and outline of a framework to both evaluate and implement fair systems. Although technical work tends to give a brief introduction to the particular definition of fairness is used, understandably, not enough reasoning is given to decide if that was a good decision. In order for roboticists to make real progress in fair robotics, we must be able to ask and decide for ourselves which definition of fairness we are implementing. 
\smallbreak
Two good thought experiments for deciding what kind of framework we might lay out are loan applications and automated package deliveries. The first takes the aforementioned definition of an algorithm as a digital robot. Loan application approval is a common example in algorithmic fairness and modern egalitarianism such as work by Meyer \cite{meyercredit}. When setting up such a system, we might refer to the various definitions laid out above and ask ourselves, "Which is most appropriate for this task?" The answer to this should be based on an interdisciplinary approach, supported by the definitions and explications laid out here. The question of an automated package delivery robot provides a more concrete example in terms of physicality, however, we are still concerned with the algorithm that directs our robot on how to deliver the packages. Borrowing the rough set-up from \cite{BRANDAO2020103259}, we can readily imagine ways in which historical bias would creep in through spatial data i.e. areas with histories of over-policing and racial policing would potentially be avoided. Here we are not directly helped by saying, ``Similar individuals should be treated similarly," as it is unclear what makes individuals similar. If we say that they are similar based on the area they live in, then we have not done anything to remove bias. Here, we might look to the Rawlsian definition to answer how to incorporate the biased data such that individuals are treated justly. It might be that the user of this package delivery service has to agree to some ranking of safety of delivery based on their location. Per Rawls, this would be fair. Another view, that of relational egalitarianism, might push us to apply more ``feedforward" morality to the issue and think about the factors that led recipients to live where they do. This might drive us to substantially devalue the biased data in favor of collecting new data or taking on a socio-spatial ambiguous navigation metric. 
\smallbreak
While these examples are simple, they help to outline the reasoning process that roboticists should adopt going forward. The problem of how to evaluate the definition of what is fair is one that should be answered not only by philosophers but also those who use fairness in their work. Although the technical problems are still at a point of vagueness due to their lack of development, looking to the future, it is clearer how we might as roboticists develop systems with fairness (and it's definition) in mind. 
\section{Conclusion}
My hope is that at this point, the implications are clear.  In the field of computer science, we must not rely on a single definition of fairness such as luck egalitarianism would have us do.  Relational egalitarianism seems more promising. This is the direction that the current \emph{similar treated similarly} seems to be headed.  However, it is also important as the designers of such algorithms that we not pass off decisions solely on society.  We must have some notion of fairness in our design that can then be leveraged to be contextualized to setting. It is not sufficient to rely solely on the vanilla definitions of fairness sourced from computer scientists. As roboticists, it is essential that we think about what fairness means as we architect a system not as a post hoc project. The support for this can be directly sourced from Elizabeth Anderson's definition. Relying only on society is the result of a distributive pattern matching. We instead must frame our project as imbuing our agents (robotic and otherwise) with some sense of virtue. This sets up the challenge as providing a methodology for robotic morals. This cannot simply be done via example but also requires a consideration of just what it means to be moral/ethical. This has begun to be tackled in works such as by \cite{DBLP:conf/robophilosophy/Kuipers16} but must become a part of our practice as a whole.

\addtolength{\textheight}{-12cm}   % This command serves to balance the column lengths

%%%%%%%%%%%%%%%%%%%%%%%%%%%%%%%%%%%%%%%%%%%%%%%%%%%%%%%%%%%%%%%%%%%%%%%%%%%%%%%%

%%%%%%%%%%%%%%%%%%%%%%%%%%%%%%%%%%%%%%%%%%%%%%%%%%%%%%%%%%%%%%%%%%%%%%%%%%%%%%%%

%%%%%%%%%%%%%%%%%%%%%%%%%%%%%%%%%%%%%%%%%%%%%%%%%%%%%%%%%%%%%%%%%%%%%%%%%%%%%%%%

\section*{ACKNOWLEDGMENT}

Many thanks to Professor Cynthia Dwork for teaching as well as feedback on this paper. Additional thanks to Lily Hu for intial discussions surrounding this work.

\bibliographystyle{IEEEtran}
\bibliography{IEEEabrv,sources}

\end{document}